\documentclass[epj]{svjour} %
\usepackage{epsfig}

\newcommand{\beq}{\begin{equation}}
\newcommand{\eeq}{\end{equation}}
\newcommand{\beqn}{\begin{eqnarray}}
\newcommand{\beqqn}{\begin{eqnarray*}}
\newcommand{\eeqn}{\end{eqnarray}}
\newcommand{\eeqqn}{\end{eqnarray*}}
\newcommand{\refeq}[1]{(\ref{#1})}

\begin{document}
\title{Target Fragmentation in $pp$, $ep$ and $\gamma p$ \\ 
Collisions at High Energies}
\author{U. D'Alesio 
\and H.J. Pirner 
}                     
\offprints{H.J. Pirner, Institut f\"ur Theoretische Physik,  
 Universit\"at Heidelberg, Philosophenweg 19, D-69120 Heidelberg
}          
\mail{umberto.dalesio@ca.infn.it}
\institute{Institut f\"ur Theoretische Physik, Universit\"at Heidelberg,
 Philosophenweg 19, 69120 Heidelberg, Germany}
\date{Received: date / Revised version: date}
%
\abstract{We calculate target fragmentation in $p p \rightarrow n X$
and $\gamma p \rightarrow n X $ reactions in the meson cloud
picture of the nucleon.
The $p p \rightarrow n X$ reaction is used to fix the
$pn\pi^+$ form factor for three different models.
We take into account the possible destruction of
the residual neutron by the projectile. Using the form factor
from the hadronic reaction we calculate photoproduction and small
$x_{\rm Bj}$ electroproduction of forward neutrons at HERA. Here the $q
\bar q$
dipoles in the photon can rescatter on the residual neutron.
In photoproduction we observe slightly less  absorption
than in the hadronic reaction.
For deep inelastic events ($Q^2>10$ GeV$^2$) screening
is weaker but still present at large $Q^2$.
The signature for this absorptive rescattering
is a shift of the  $d\sigma/dE_n$ distribution to higher neutron
energies for photofragmentation.
} 
\maketitle
%


\section{Introduction}

In the last two decades the study of total inclusive deep inelastic
scattering
(DIS) processes has allowed to extract important information on the
structure of hadrons. Parton distributions have been
determined and scaling violations have been tested to a high level of
accuracy. The QCD improved parton model has been shown to be very
reliable in the presence of a hard scale.  

On the other hand semi-inclusive reactions with electromagnetic
probes are still less explored. They can improve
our knowledge on the inner hadronic structure. \\
Through the study of new
observables characterizing these processes we may ask more detailed
questions about the proton. In this case the application of
perturbative QCD (pQCD) has been restricted to high $p_t$
events, and experiments have been
analyzed in terms of parton distributions and fragmentation
functions.

With the HERA collider target fragmentation  can be
studied in a much cleaner way than with fixed target experiments.  
New interest has been triggered 
by measurements on leading neutron production performed 
by the  ZEUS and H1  collaborations at the
electron-proton collider  \cite{Ze,H1}.
These data are currently  analyzed in terms of hadronic degrees of
freedom, i.e. studying the virtual pion flux in the nucleon 
\cite{HLNSS,KPP}. 
Thus one hopes to extract information
about the pion structure function at very small $x$, not reachable in 
Drell-Yan experiments.\\
In the late fifties Chew and Low already suggested the idea of using 
pions from the pion cloud of the proton as targets to get information 
on the interaction of different projectiles with pions \cite{CL}.\\
Novel theoretical tools 
to include forward leading particle production in the
framework of standard pQCD have been developed by
Trentadue, Veneziano and Graudenz \cite{TrVe,Gr}, who introduced
a new set of nonperturbative distributions (fracture 
functions) which allow to absorb collinear singularities at leading
order in the QCD coupling constant.

If we look at  $\gamma^* p$  reactions in the cm-system, we can consider
the
incoming photon as a $q \bar q$ state \cite{NZ}
that interacts with a proton
made up of two color neutral components, one of which is the
final state neutron. This involves quite different aspects of the 
nucleon than those investigated in deep inelastic inclusive
scattering.
Especially long range properties  of the baryon can be analyzed
in such a process. At which length 
does the string connecting a quark to the residual diquark break
and produce
two colour neutral objects, a meson and a nucleon? This 
question is especially important for  nuclear physics,
as one wants to know what amount of the nucleon-nucleon
interaction
is describable in terms of  meson exchange forces in the nucleus.

From the high-energy point of view 
the determination of the pion structure function is central. 
The common interpretation of the $\gamma^* p \rightarrow nX$
experiment 
relies on the application of the meson cloud model of the nucleon 
\cite{Zo,HSS,KFS} 
together with the factorization hypothesis that allows to separate the
reaction into two steps: the fragmentation process and the interaction
\cite{Su,Bi}.
It   assumes that   
the fragmentation process is universal, i.e.  
independent of the projectile
which initiates fragmentation.
In the following  we 
will examine the validity of this factorization hypothesis
carefully.

\begin{figure}[htb]
\center
\mbox{\epsfig{file=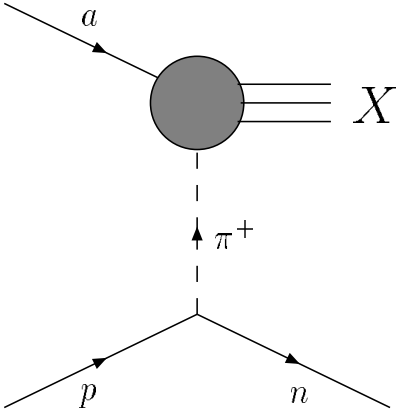,width=6cm}}
\caption{\label{opedia}
{ Picture of semi-inclusive neutron production according to the 
one-pion-exchange model.}}
\end{figure}

Let us consider the
generic reaction with an incoming projectile
$a$ leading to neutron production i.e. 
$ a p \rightarrow n X $  (see fig.~\ref{opedia}).
In the one-pion-exchange model the differential cross section
is given by the product of the pion flux factor
times the total $a\pi$ cross section:
\beq
\label{fact}
\frac{d\sigma^{a p \rightarrow n X }}{dzdp_t^2} = 
        F_{n\pi}(z,p_t)\,\sigma^{a\pi }_{\rm tot}(s') \,.
\eeq
The flux factor $F_{n\pi }(z,p_t)$ gives the probability for 
the splitting of a proton into a pion-neutron system.
It depends on  the longitudinal momentum fraction $z$ carried by the
detected neutron and  its transverse momentum\footnote{We
use $p_t = |{\bf p}_t|$.} $ p_t$. The total
cross section $\sigma^{a\pi }_{\rm tot}(s')$ is  a function of
$s'$ the
center of mass sub-energy squared in the $a\, \pi^+$ interaction.

The usual procedure  \cite{HLNSS,KPP} is  to fix the  parameters
entering the flux factor
from the data of leading neutron production in proton-proton ($a=p$)
collisions. Then assuming universality, one  applies the same
equation
\refeq{fact} to virtual photon scattering ($a=\gamma^*$) in DIS.
Measuring the differential cross section $ \gamma^* p \rightarrow n X $  
one extracts
the pion structure function, since
$
\sigma_{\rm tot}^{\gamma ^*\pi^+}$ is proportional to $F_2^{\pi^+}$. 
Obviously the factorization
hypothesis plays a crucial role in this method.

Indeed early works \cite{MTU,Az,ZS} on absorption effects for pion-exchange 
mechanism in hadronic inclusive reactions indicate a suppression factor 
of the order of 50-70\%.
Recently a first and detailed study of absorptive corrections 
in the Regge formalism
\cite {NSZ} has appeared.
Many of our conclusions agree with the ones reached in ref.~\cite {NSZ}. 
These absorptive effects
depend on the projectile and are a source of factorization
breaking.
They can be  comparable or even  more important than other background
contributions, already estimated, reducing the accuracy of
$F_2^{\pi^+}$ measurements.  
We will investigate  the relevance of absorptive corrections
in detail in order to  understand 
the one-pion-exchange mechanism and the extraction of the pion structure
function. 
We apply high-energy Glauber theory to
calculate and compare the screening corrections in 
leading neutron production for  $p\, p$ and $\gamma^{(*)}\, p$ reactions.
We especially search for differences between photoproduction and deep
inelastic scattering in neutron fragmentation.
To this end we follow the $Q^2$-evolution 
of the fracture functions from a kinematic region where
the perturbative evolution of \cite {TrVe,Gr} is not yet
applicable to large $Q^2$. 
We find similarly to ref.~\cite{NSZ} that absorptive corrections for 
highly virtual photons do not vanish, as indicated by data \cite{Ze,JaNu}. 
However this effect is smaller than what has been found for proton-proton 
interactions. This dependence on the projectile can be a source of 
factorization breaking. 
With respect to ref.~\cite{NSZ}, we extend this analysis to real photons 
($Q^2=0$)  and we find that the size of rescattering corrections is 
comparable to that of hadronic reactions and bigger than that of virtual 
photons.

The outline of the paper is as follows. In section 2 we describe the
meson cloud model, in section 3 we calculate the absorptive corrections
to $p p \rightarrow n X$. Section 4 is devoted to virtual and real  
photoproduction of forward neutrons. Section 5 closes with a summary and a 
discussion.

\section{Meson Cloud Model}

In this section we review briefly the main features of the meson cloud
model (MCM).\\
In the MCM a proton is viewed as a bare proton surrounded by a virtual
meson cloud,
\beqn
\label{MCM}
 |p\uparrow\rangle & = & \sqrt{S}\,\Big\{ |p_0\uparrow\rangle\nonumber\\   
&&          + \sum_{\lambda\lambda'}\sum_{BM}\int\!
             dzd^2{\bf p}_t\,\phi_{BM}^{\lambda\lambda'}(z,{\bf p}_t)
             |B,M;z,{\bf p}_t\rangle\Big\}\nonumber \\
 & = & \sqrt{S}\,\Big\{ |p_0\uparrow\rangle 
        +\sum_{\lambda\lambda'}\int\!
     dzd^2{\bf p}_t\,\phi_{N\pi}^{\lambda\lambda'}(z,{\bf p}_t)\\
&& \mbox{}\times  \Big[\sqrt{\frac{1}{3}}|p,\pi^0;z,{\bf p}_t\rangle
        + \sqrt{\frac{2}{3}}|n,\pi^+;z,{\bf p}_t\rangle  
        \Big] + \cdots\Big\}\,,\nonumber
\eeqn
where $\phi_{BM}^{\lambda\lambda'}(z,{\bf p}_t)$ is the probability
amplitude to
find, inside a proton with spin up, a baryon $B$ with longitudinal
momentum fraction
$z$, transverse momentum ${\bf p}_t$ and helicity $\lambda$ and a meson
$M$, with longitudinal momentum fraction $1-z$, transverse momentum
$-{\bf p}_t$ and helicity $\lambda'$. We restrict ourselves
to the first contributions of this expansion in terms of Fock states. 
$\sqrt{S}$ is the renormalization constant, which is fixed by
$\langle p|p \rangle = 1$ and gives the amplitude for the bare proton.

In the light-cone approach the amplitudes $\phi_{N\pi}$, for a proton with
spin $+1/2$, read \cite{HSS} 
\beqn
\label{phi}
\phi_{N\pi}^{1/2,0}(z,{\bf p}_t) & = & 
          \frac{\sqrt{3}g_0}{4\pi\sqrt{\pi}}\frac{1}{\sqrt{z^2(1-z)}}
          \frac{m_N(z-1)}{M_{N\pi}^2-m_N^2}\nonumber\\
\phi_{N\pi}^{-1/2,0}(z,{\bf p}_t) & = & 
          \frac{\sqrt{3}g_0}{4\pi\sqrt{\pi}}\frac{1}{\sqrt{z^2(1-z)}}
          \frac{|{\bf p}_t|e^{-i\varphi}}{M_{N\pi}^2-m_N^2}\,,
\eeqn
where $M_{N\pi}^2$ is the invariant mass of the pion-nucleon system,
given by
\[
M_{N\pi}^2 = \frac{m_N^2+p_t^2}{z} + \frac{m_\pi^2+p_t^2}{1-z}\,,
\]
$m_N$ and $m_\pi$ are the nucleon and the pion masses; $g_0$ is the bare 
pion-nucleon coupling constant and $\varphi$ is the azimuthal angle
in the transverse plane. Including the renormalization factor $\sqrt{S}$
from eq.~\refeq{MCM} 
we get the renormalized effective coupling $g = \sqrt{S}g_0$
\cite {KFS}, which can
be extracted from low-energy data: we use $g^2/4\pi = 13.75$.

Because of the extended nature of the
hadrons involved, the interaction amplitudes in eq.~\refeq{phi} 
have to be modified by including  a phenomenological $\pi NN$ form factor. 
It is important to stress here that while the vertex is derived
from an effective meson-nucleon Lagrangian, the form factor 
is introduced {\it ad hoc}.
In order to parametrize the form factor we need to
introduce the momentum transfer $t$ which 
can be expressed as follows
\beqn
t = (p_N-p_N')^2 & = & 
        -\frac{1}{z}\left[p_t^2+(1-z)^2 m_N^2\right]\nonumber\\
& = &   (1-z)(m_N^2-M_{N\pi}^2) + m_\pi^2\,. \nonumber
\eeqn

Different models and parametrizations are available in the literature.
In the following we compare the results obtained using 
the light-cone approach and the covariant approach, 
the last one with inclusion of reggeization. 
Besides these two form factors we consider a $\pi NN$ form factor
extracted from Skyrme-type models \cite{HoMa,FrSc}.
This form factor leaves low momentum transfers essentially unaffected
while suppressing the high momentum region strongly. The three models
are: 
\begin{itemize}
\item light-cone $\pi NN$ form factor
\beqn
\label{lcff}
G(z,p_t) & = & {\rm exp}[R_{lc}^2(m_N^2-M_{N\pi}^2)] \nonumber\\
         & = & {\rm exp}[R_{lc}^2(t-m_\pi^2)/(1-z)]
\eeqn
\item covariant $\pi NN$ form factor
\beq
\label{covff}
G(z,p_t)  =  {\rm exp}[R_{c}^2(t-m_\pi^2)]
\eeq
\item Skyrme model $\pi NN$ form factor
\beq
\label{Sky}
G(z,p_t)  =  {\rm exp}[R_{S}^2\,t] g(t)
\eeq
where $g(t)$ is a rational function of $t$ given in the appendix.
\end{itemize}
The amplitudes $\phi^{\lambda\lambda '}$ must be changed according to
$
\phi^{\lambda\lambda '} \rightarrow \phi^{\lambda\lambda '}\,G(z,p_t)\,.
$
The flux in eq.~\refeq{fact}, for proton fragmentation into a neutron can
then be calculated as
\beq
\label{flux}
F_{n\pi}(z,p_t) = \frac{2}{3}\pi\sum_{\lambda\lambda'}
        |\phi_{N\pi}^{\lambda\lambda'}(z,{\bf p}_t)|^2|G(z,p_t)|^2\,,
\eeq
where $2/3$ is the isospin factor and
the azimuthal angle in the transverse plane
has been integrated out.
The reggeization of the pion (relevant for $z\rightarrow 1$)
is included in the covariant and Skyrme approaches by the further change
\[
\phi^{\lambda\lambda '} \rightarrow 
        \phi^{\lambda\lambda '}(1-z)^{-\alpha_\pi(t)}\,,
\]
where $\alpha_\pi(t) = \alpha_\pi(0) + \alpha_\pi't$ is the pion 
Regge-trajectory, with $\alpha_{\pi}(0) = 0$ and $\alpha_{\pi}' \simeq 1
\,{\rm GeV}^{-2}$.
The light-cone form factor contains the decrease of the cross section
for $z\rightarrow 1$ already in the exponential. 
Its form, however, is a  crude approximation and therefore we do not
expect the light-cone form factor to be adequate for extremely large $z$.

In the following section we consider
$pp$ collisions in order to fix the  parameters 
$R_{lc}$ and $R_c$ appearing in the previous
equations\footnote{For the Skyrme form factor we use the value
fitted in \cite{HoMa}: $R_s^2 = 0.031/m_\pi^2$.}.

\section{Estimate of absorptive corrections in $pp\rightarrow nX$}

We consider  target fragmentation reactions as  stripping 
reactions in the cm-system where the projectile proton
strips a $\pi^+$ from the 
target proton leaving behind a neutron.
The projectile proton smashes the pion into pieces, while 
the neutron remains intact as a spectator. 
Any  additional interactions
like $\Delta-$production or $\rho$-exchange may spoil this
simple picture and
reduce the accuracy of the determination of the $\pi pn$ vertex. This has
been
studied and the amount of such a  background is estimated to be around
20\% \cite{HLNSS,KPP}. 
We will neglect these processes in our calculation, but 
we model them rescaling the one-pion-exchange cross section by 1.2.

We are fully aware that this rescaling procedure is a poor simplification
of the background contributions. A recent work \cite{NSSS}
shows that an improved treatment of the background is important to get the 
right neutron distribution at very high $z$ values. 
Nevertheless we will 
concentrate mainly on the region of $p_t=0$ and $0.7<z<0.9$, where our  
simple rescaling gives reasonable results.

\begin{figure*}[hbt]
\center
\mbox{\epsfig{file=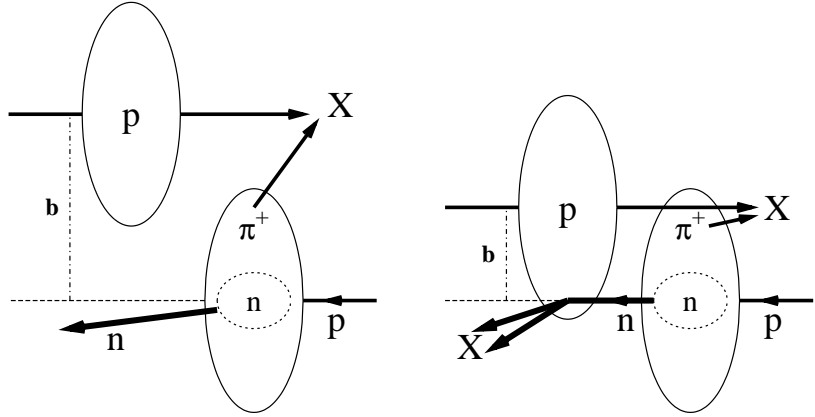,width=11cm}}
\caption{\label{glaureact}
{ Picture of the collision for two different impact parameters. 
On the left 
we show a peripheral collision (large ${\bf b}$) where the pion 
is stripped and the neutron acts as a spectator. On the right we show a
more central collision where the projectile proton  can
destroy the neutron through rescattering.}}
\end{figure*}

The invariant differential cross section for the one-pion-exchange
mechanism is (in light-cone approach\footnote{Reggeized covariant
expression can be obtained multiplying eq.~\refeq{opepp} by
$(1-z)^{-2\alpha_\pi(t)}$ and taking $G(z,p_t)$ from eq.~\refeq{covff}.}) 
\beqn
\label{opepp}
E_n\frac{d^3\sigma}{d^3 {\bf p}_n} &=& 
        \frac{z}{\pi}\frac{d\sigma}{dzdp_t^2} \\   
        &=& \frac{2g^2}{16\pi^2}\frac{1}{z(1-z)} 
        \frac{m_N^2(1-z)^2+p_t^2}{(M_{N\pi}^2-m_N^2)^2}
         |G(z,p_t)|^2 \sigma_{\rm tot}^{p\pi }\,.\nonumber
\eeqn
This picture is reliable when the pion and the neutron in the {\it target\/}
proton are well separated, i.e.  at  large $z$ and 
large impact parameter. For small impact parameters and intermediate $z$
values  it  
must be extended to allow 
the scattering of the projectile on the neutron and
the consequent screening effect (see fig.~\ref{glaureact}).

To be more precise we distinguish in our
notation between the target 
($p_{(T)}$) and projectile ($p_{(P)}$) proton:
\[
p_{(P)}\, p_{(T)} \rightarrow n X\,.
\]
We employ the high-energy Glauber 
approximation to multiple scattering which
has been  used for target
fragmentation in heavy ion collisions in ref. \cite {AMHu}.

A more formal derivation of absorptive corrections can be found in 
\cite{NSZ}, where Regge calculus and generalized AGK cutting rules 
\cite{CKT} are applied. 
Our approach can be traced back to that of \cite{NSZ} and viewed as 
a simplification of it in the eikonal approximation.

We treat the target proton 
as a pion-neutron system ($\phi_0$) undergoing a  
transition to an excited state
($\phi_\alpha$).
The cross section for this process can be expressed as
\beqn
\label{pp1step}
\sigma(\phi_0\rightarrow\phi_\alpha)  &=& 
        \int\!d^2{\bf b}\,\sigma_{0\rightarrow\alpha}({\bf b})
        \nonumber\\
        &=& 
        \int\!d^2{\bf b}\,\left|\langle\phi_\alpha [1-S^{p\pi} 
        S^{pn}]|\phi_0\rangle\right|^2\,,
\eeqn
where ${\bf b}$ is the impact parameter and $S^{ab}$ are the interaction
operators (see below).
We assume that the proton state $\phi_0$ can be factorized into a 
system of a pion and a neutron 
\[
|\phi_0\rangle \equiv |\psi_0^{ \rm sp}\pi_0n_0\rangle = 
|\psi_0^{ \rm sp}\rangle|\pi_0\rangle|n_0\rangle\,,
\]
with $\psi_0^{\rm sp}$ the spatial component, specified to keep spatial 
and intrinsic degrees of freedom separated. 
Similarly the excited state with an undisturbated neutron 
has the form:
\[
|\phi_\alpha\rangle \equiv |\psi_j^{\rm sp}\pi_{\alpha'} n_0\rangle =
|\psi_j^{\rm sp}\rangle|\pi_{\alpha'}\rangle|n_0\rangle\,,
\]
where we have introduced an extra index ($j$) to take into account all
possible spatial configurations for this excited state;
$|\pi_{\alpha'}\rangle$ is an arbitrary  state with the same quantum
numbers as the pion. 
To get the total cross section we sum now over all spatial configurations
and over all $|\pi_{\alpha'}\rangle$ states. We apply
the closure relation and exclude the elastic
contribution ($|\pi_0\rangle$)
\beqn
\label{sigmab}
\sigma({\bf b}) & = & \sum_{\alpha} 
        \sigma_{0\rightarrow\alpha}({\bf b})
        =  \sum_{\alpha'\neq 0}\sum_j\Big|
        \int\!d^3y_n d^3y_\pi \nonumber\\
& & \hspace{.5cm} \psi_j^*(y_n,y_\pi)\psi_0(y_n,y_\pi)
        S_{\alpha' 0}^{p\pi}({\bf b}-{\bf s}_\pi)
        S_{00}^{pn}({\bf b}-{\bf s}_n)\Big|^2 \nonumber\\
& = &   \sum_{\alpha'\neq 0} \int\!d^3y_n d^3y_\pi 
        |\psi_0(y_n,y_\pi)|^2 \nonumber\\
&&\mbox{}
        \hspace{.5cm}\times S_{\alpha' 0}^{p\pi}({\bf b}-{\bf s}_\pi)
        S_{0\alpha'}^{\dag p\pi}({\bf b}-{\bf s}_\pi)
        |S_{00}^{pn}({\bf b}-{\bf s}_n)|^2 \nonumber\\
& = &   \int\!d^3y_n d^3y_\pi|\psi_0(y_n,y_\pi)|^2\nonumber\\
&&\mbox{} \hspace{.5cm}\times
        \left[1-|S_{00}^{p\pi}({\bf b}-{\bf s}_\pi)|^2\right]
        |S_{00}^{pn}({\bf b}-{\bf s}_n)|^2 \,,
\eeqn
where $y_n \equiv ({\bf s}_n,z_n)$, 
$y_\pi \equiv ({\bf s}_\pi,z_\pi)$, ${\bf s}_{\pi}$
and ${\bf s}_{n}$ are the
coordinates of the pion and the neutron in the impact parameter plane;
$z_n$ and $z_\pi$ are their longitudinal momentum fractions. Also
\beqn
\label{prof}
1-|S_{00}^{p\pi}|^2  &=&  1 - |1-\Gamma^{p\pi}|^2 \simeq 2Re\Gamma^{p\pi}
\nonumber\\
|S_{00}^{pn}|^2  &=&  |1-\Gamma^{pn}|^2 \simeq 1-2Re\Gamma^{pn}\,,
\eeqn
where the
profile functions $\Gamma$ describe the 
respective two-body scatterings and are related
to the scattering amplitudes in momentum space by Fourier transformation 
\beq
f^{ij}({\bf q}) = \frac{ip_{\rm cm}}{2\pi}\int\!d^2{\bf b}\, e^{i{\bf
q\cdot
b}}\,\Gamma^{ij}({\bf b})\,.
\eeq
Let us now consider the density distribution $|\psi_0(y_n,y_\pi)|^2$. We
parametrize this density starting from the probability density to find a
pion and a neutron at a certain transverse separation ${\bf b}_{rel} =
{\bf s}_n-{\bf s}_\pi$, 
imposing the center-of-mass constraint in ${\bf b}$-space
and the longitudinal momentum conservation: 
${\bf b}_{cm}  = z_n{\bf s}_n + z_\pi{\bf s}_{\pi} = 0 $, 
$z_n+z_\pi=1$, 
\beq
\label{psi}
|\psi_0(y_n,y_\pi)|^2 \equiv  \rho_{n\pi}(z_n,{\bf b}_{rel})
\delta^2(z_n{\bf s}_n + z_\pi{\bf s}_{\pi})\delta(z_n+z_\pi-1)\,.
\eeq
Inserting eqs.~\refeq{prof} and \refeq{psi} into eq.~\refeq{sigmab} and
carrying out 
the 3-dimensional integration over  $y_\pi$, 
we get for the semi-inclusive cross section
\beq
\label{dsdz}
\frac{d\sigma^{pp\rightarrow nX}}{dz}  =  \int\!d^2{\bf
        b}\,\frac{d\sigma({\bf b})}{dz}\,, \hspace{1cm}{\rm with}
\eeq
\beqn
\label{ppglau}
\frac{d\sigma({\bf b})}{dz} & = & \int\!d^2{\bf s}_n\frac{1}{(1-z)^2}\,
        \rho_{n\pi}(z,{\bf b}_{rel})\nonumber\\ 
&&\mbox{} \times 2Re\Gamma^{p\pi}({\bf b}-{\bf s}_\pi)
        \left[1-2Re\Gamma^{pn}({\bf b}-{\bf s}_n)\right]\nonumber\\
& = & \int\!d^2{\bf b}_{rel} \rho_{n\pi}(z,{\bf b}_{rel})\,
  2Re\Gamma^{p\pi}({\bf b}+z{\bf b}_{rel})\nonumber\\
&&\mbox{}\times
        \left[1-2Re\Gamma^{pn}({\bf b}-(1-z){\bf b}_{rel})\right]\,,
\eeqn
where we have restored the previous notation, $z_n \equiv z$ and 
replaced the pion and neutron coordinates by 
\[
{\bf b}_{rel} = -\frac{{\bf s}_\pi}{z} = \frac{{\bf s}_n}{1-z} \,.
\]

\begin{figure*}[h!tb]
\center
\mbox{\epsfig{file=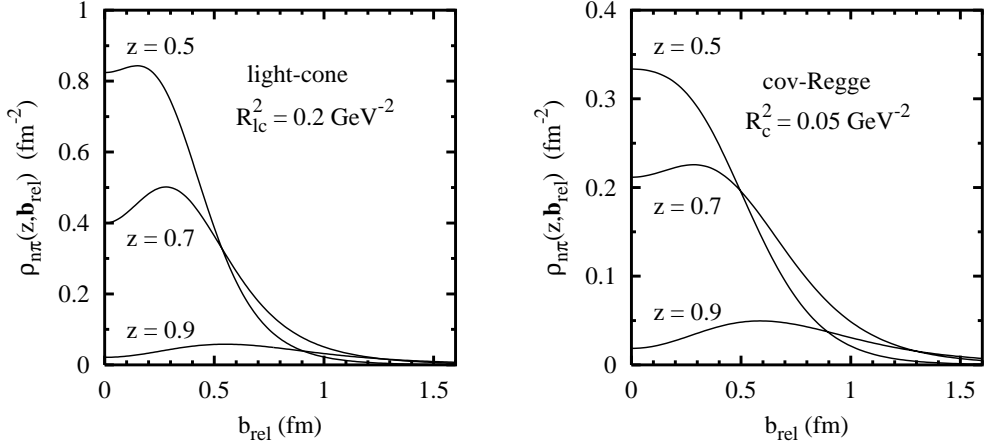,width=14cm}}
\caption{\label{dens}
{ Pion-neutron densities in the transverse  plane 
for various $z$ values in light-cone (left) and covariant
approach (right).}}
\end{figure*}

\subsection{Pion-neutron density  and profile functions in 
${\bf b}$-space}

In order to extract quantitative information from
eq.~\refeq{ppglau} we evaluate the probability density to
find a pion and neutron at a certain distance inside the parent proton.
The main idea is to start with the amplitudes in momentum space for
the splitting of a proton into a pion-neutron system (see
eqs.~\refeq{phi}) and then calculate  the
Fourier transform in two dimensions with respect to the transverse
momentum. \\
We obtain the amplitudes in ${\bf b}$-space (we keep $z$ fixed)
\[
\psi^i_{n\pi}(z,{\bf b}_{rel}) = \frac{1}{2\pi}\int\!d^2{\bf p}_t 
        e^{i{\bf b}_{rel}\cdot{\bf p}_t}\phi^i_{n\pi}(z,{\bf p}_t) \,,
\]
where by $\phi^i_{n\pi}$ we mean
$\sqrt{2/3}\phi^{\lambda\lambda'}_{N\pi}$. 
From this we obtain the probability density to find a neutron  and a 
pion respectively with longitudinal momentum $z$ and $1-z$ and relative
transverse separation ${\bf b}_{rel}$:
\[
\rho_{n\pi}(z,{\bf b}_{rel}) = \sum_i|\psi^i_{n\pi}(z,{\bf b}_{rel})|^2\,.
\]
In fig.~\ref{dens} we show the behaviour of these densities at different
values of $z$ for the light-cone and the covariant approach.

Concerning the profile functions we start from scattering amplitudes with
gaussian shapes,
\beq
f^{ab}({\bf q}) = \frac{ip_{\rm cm}}{4\pi}\sigma_{\rm tot}^{ab} 
\,{\rm exp}\left[-\frac{{\bf q}^2}{2\Lambda^2_{ab}}\right]
\eeq
and by Fourier transformation  we get
\beq
\Gamma^{ab}({\bf b}) = \frac{1}{4\pi}\sigma_{\rm tot}^{ab}\Lambda_{ab}^2
\,{\rm exp}\left[-\frac{{\bf b}^2\Lambda^2_{ab}}{2}\right]\,,
\eeq
where we have considered purely imaginary amplitudes as we are interested
in the high-energy regime.

\subsection{Cross section for $p p \rightarrow n X$ }

The gaussian dependence of the profile functions allows us to perform
the ${\bf b}$-integration  in eq.~\refeq{dsdz} analytically, we have then
\beqn
\label{dsdzfin}
\frac{d\sigma^{pp\rightarrow nX}}{dz} &=& \int\!d^2{\bf b}_{rel}
        \rho_{n\pi}(z,{\bf b}_{rel})\,\sigma_{\rm tot}^{p\pi^+}\\
&&      \mbox{}\times
        \Big\{1-\Lambda_{\rm eff}^2\frac{\sigma_{\rm tot}^{pn}}{2\pi}\,
        {\rm exp}\Big[-\frac{\Lambda_{\rm eff}^2{\bf
        b}_{rel}^2}{2}\Big]\Big\}\,,\nonumber
\eeqn
with 
\beq
\Lambda_{\rm eff}^2 =
\frac{\Lambda_{p\pi}^2\Lambda_{pn}^2}{\Lambda_{p\pi}^2+\Lambda_{pn}^2}
\hspace{1cm} [\Lambda_{p\pi}^2 \approx \Lambda_{pn}^2\approx 
0.08\,{\rm GeV}^{2}]\,.
\eeq
Before giving explicit results from eq.~\refeq{dsdzfin}, let us try
to interprete the physical picture emerging from it.\\
The first term alone is nothing else than
the standard expression according to the factorization hypothesis,
eq.~\refeq{fact}, i.e. the stripping of the pion cloud inside the
target proton.
The second term represents the screening correction
which is the most interesting result of our
calculation. The Born fragmentation cross section
is multiplied  by the probability that the projectile proton does
not destroy the neutron component of the target proton. 
The screening factor
changes the simple factorization picture.

Assuming  that the
final state interaction does not modify the transverse 
momentum distribution of the fragments,
we can calculate the invariant differential cross
section  
$E_n\frac{d^3\sigma}{d^3{\bf p}_n}$  by multiplying
the differential cross sections 
for the longitudinal distributions with  the transverse probability
distribution for the fragmentation process. 
This method seems to work quite well in nuclear target fragmentation
\cite {AMHu}. We get 
\beqn
\label{d3csgl}
E_n \frac{d^3\sigma^{pp\rightarrow nX}}{d^3{\bf p}_n}
& = & \frac{z}{\pi}\frac{1}{N(z)}\frac{dN(z,p_t)}{dp_t^2}\,
\frac{d\sigma^{pp\rightarrow nX}}{dz} \,,
\eeqn 
where $d\sigma/dz$ is given by eq.~\refeq{dsdzfin}. The normalized
fraction of
fragmentation processes in the interval
$p_t$, $p_t+dp_t$ is obtained from the normalized pion flux factor:
\beq
\label{normflux}
\frac{1}{N(z)}\frac{dN(z,p_t)}{dp^2_t}= \frac{F_{n\pi}(z,p_t)}
        {\int\!dp_t^2 F_{n\pi}(z,p_t)} \,.
\eeq

\begin{figure}[!thb]
\center
\mbox{\epsfig{file=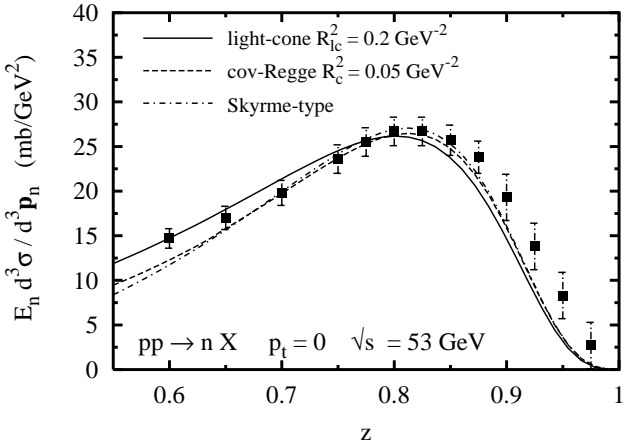,width=8.5cm}}
\caption{\label{dsd3p0} 
{ Invariant differential cross sections for neutron production 
at $p_t = 0$ calculated according to eq.~\refeq{d3csgl} for three
different form factors. Data points are
from ref.~\cite{FlMo}.
}}
\end{figure}

\begin{figure}[!thb]
\center
\mbox{\epsfig{file=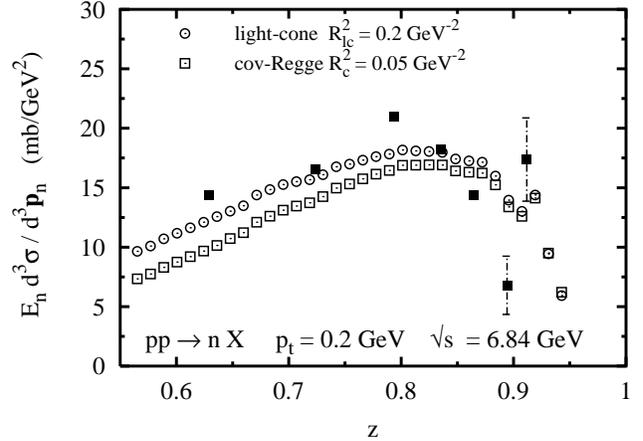,width=8.5cm}} 
\caption{\label{dsd3p02} 
{ Invariant differential cross sections for neutron production 
at $p_t = 0.2$ GeV calculated according to eq.~\refeq{d3csgl} for two
different form factors. Skyrme-type approach, not displayed, gives almost 
the same result as the covariant one.  
Points are low-energy $\sqrt s = 6.84$ GeV bubble chamber data, 
ref.~\cite{Bl}.
}}
\end{figure}

Definitely the experimental  cross sections for  $p_t=0$ \cite {FlMo,Bl}
have the most pronounced
shape. Here one can really see a fragmentation peak which seems
to be superimposed on  some background. 
We estimate additional contributions\footnote{These  may come from
resonance excitations like $p\rightarrow \pi^+\Delta^0$ and
$p\rightarrow\pi^+ N^{*0}$, which decay into neutrons.} to have a similar 
shape as the calculated cross section and lead to a 20\% 
correction. 
Thus we scale
our result for the one-pion-exchange contribution by a factor 1.2.
A model calculation of background processes can be found in ref.
\cite {NSSS}.
For larger $p_t$ values 
the fragmentation cross sections become rapidly flatter and 
smaller and the background increases. 
Therefore we fit the radius parameters to the $p_t=0$ 
data.
In fig.~\ref{dsd3p0} we show the semi-inclusive cross section for 
$p p \rightarrow n X$ as a function of the longitudinal momentum
fraction $z$ of the neutron together with the experimental data from
ref.~\cite {FlMo}. 
For the total cross sections $\sigma^{p\pi^+}_{\rm tot}$ 
and $\sigma^{pn}_{\rm tot\/}$ we 
adopt the fits\footnote{In fig.~\ref{dsd3p02} 
for $\sigma^{p\pi^+}_{\rm tot\/}$ we take
directly the scattering data at small energies from the particle data
group \cite{PDG}.} performed in ref.~\cite{DoLa}. 
We adjust the radius parameters $R_{lc}$
in the light-cone and $R_c$
in the covariant form factor to the
data and  find a reasonable agreement with the following values: 
$R_{lc}^2 = 0.2 \,{\rm GeV}^{-2}$ and
$R_{c}^2 = 0.05 \,{\rm GeV}^{-2}$. 
An important feature of the
screening correction is that its inclusion reduces
the radius parameters so extracted. 
The shape of the Skyrme form
factor does not deviate appreciably from the covariant form factor,
because even at smaller values of $z$ the 
momentum transfers $|t|$ in fig.~\ref{dsd3p0}
are not large. 
One should also keep in mind that the reggeized pion has a variable
spin different from zero, when $t$ is different from the pole value.
The coupling of virtual pions to the nucleon does not have to be identical
to that imployed 
in one-pion-exchange potentials.

The differences of the light-cone \,and 
covariant\, approaches are more 
pronounced for very large $z$ and at $p_t\neq 0\,$. 
Here the light-cone form factor vanishes
faster with $z$ compared to the fall off in
the reggeized covariant form factor. 
In fig.~\ref{dsd3p02}  we compare  the light-cone and the 
covariant-Regge  models with 
the invariant differential cross
sections measured at $p_t=0.2$ GeV \cite {Bl}. The theory does not
agree very well with these data. 
In fact these data are taken at $\sqrt s =6.84$ GeV,
where Feynman scaling is not expected to hold. 
One sees for the low-energy data that near $z=0.9$ the 
$\pi^+ p $ resonances  affect the cross section. 
There are also data at larger energies \cite {En},
but these do not cover simultaneously the large $z$ and small $p_t$
range.

\begin{figure}[!thb]
\center
\mbox{\epsfig{file=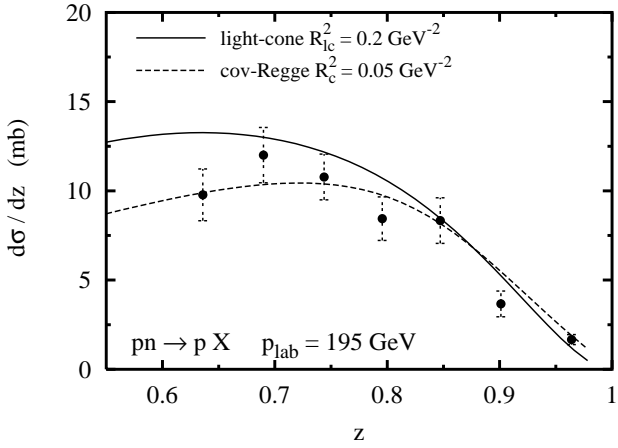,width=8.5cm}} 
\caption{\label{dsdzpn} 
{Inclusive cross section for the $pn\rightarrow pX$ process at 
$p_{\rm lab} = 195$ GeV for $|t|< 1.4$ GeV$^2$, see text.  
Data points are from \cite{Eis}.
}}
\end{figure}

The same calculation performed for the $pp\rightarrow nX$ reaction can be 
applied to $pn\rightarrow pX$, where the same $pn\pi$ vertex factor 
enters. The only changes to be done are $\sigma^{p\pi^-}$ for
$\sigma^{p\pi^+}$ and $\sigma^{pp}$ for $\sigma^{pn}$ in 
eq.\refeq{dsdzfin}. In fig.\ref{dsdzpn} we show the $p_t^2$-integrated 
cross section for the $pn\rightarrow pX$ in light-cone and 
covariant-Regge pictures with the data from 
\cite{Eis}. Again a reasonable agreement is found with the values of the
radius parameters fitted on the leading neutron data at $p_t = 0$ 
(fig.\ref{dsd3p0}).

\begin{figure}[!bht]
\center
\mbox{\epsfig{file=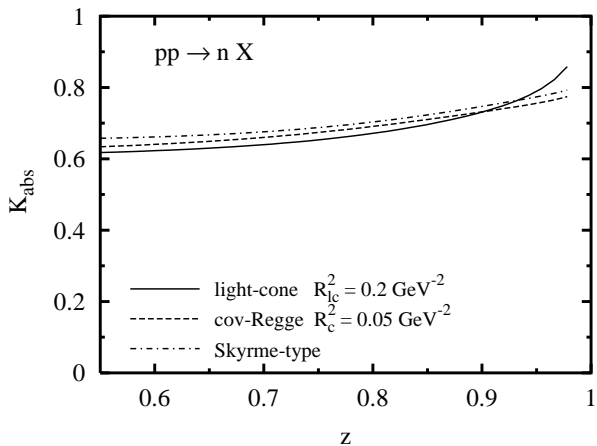,width=8.5cm}} 
\caption{\label{kabspp} 
{ Absorptive $K$-factors for neutron production in proton-proton
collisions for three different form factors.
}}
\end{figure}

The effect of screening is shown clearly in fig.~\ref{kabspp}, 
where we plot the $K$-factor, i.e. the 
ratio of the differential cross section with and without absorptive
corrections for the fitted values of the radii. 
The individual models differ slightly in the 
$z <0.8$ region where there is a sizeable, $\ge 30 \%$, screening. 
The light-cone form factor leads to a slightly smaller
$K$-factor. 
For very large $z$ values large transverse separations
between the neutron and the pion dominate \cite {NSZ}
(see fig.~\ref{dens}), 
thus the projectile
proton misses the neutron for almost any finite impact parameter
and the $K$-factor approaches unity.

A comparison of the data with a calculation neglecting the screening
mechanism would give larger radii. Such a fit can represent the
data down to lower $z$ 
values like $z=0.65$. 
In general the amount of absorption can be reduced if one uses a larger
cut-off radius in the pion form factor and additional exchanges
 \cite {NSSS}. An understanding of the trade-off between these
two effects necessitates that they are considered together.

\section {Cross sections for $\gamma^* p \rightarrow n X$ and 
$\gamma p\rightarrow n X$}

In this section we consider the interaction of an initial virtual or real 
photon with a proton leading to neutron target fragments. 
Before entering the detailed calculation of the semi-inclusive reaction it
is worthwhile to make some remarks on inclusive photon-nucleon 
interaction in context with our $\pi NN$ form factor.
Recently \cite {FrSc} an estimate of 
the antiquark distributions in the nucleon 
has been performed with 
the help  of the Sullivan formula, cf.~eq.~\refeq{fact}.
The calculation shows that for the  Skyrme form factor  used above the MCM
prediction   
exhausts or only slightly exceeds the light sea quark distributions $\bar
u$ and $\bar d$
in the proton at $Q^2=1.2\, {\rm GeV}^2$. 
This is a large improvement with respect to harder form factors often used
before, which strongly overestimate this contribution. Thus the problems
discussed e.g. in ref. \cite {KFS} 
seem to be well taken care of. Two comments are in place:
perhaps the $F_2$ calculations are not valid because of the 
interference between photon-nucleon and photon-pion inelastic
interactions, where
the target-like  slow fragments 
of the struck sub-hadron  and the residual hadron
interact \cite{NSZ}. Since both are slow
and strongly interacting there is no reason that this final state interaction
is negligible. 
A second remark about the pion pole calculation is
the possible contribution of diffractive excitation of the proton feeding the
neutron channel. This problem has been investigated in \cite{HNSSZ}. 
Preliminary experiments \cite{JaNu} rather 
point to the fact that for $z<0.8$ there is a sizable
contribution of $\pi_0$-exchange to forward proton production \cite{SNS}.

A new approach \cite{NSSS} has been recently applied 
to describe in a consistent and unified framework proton fragmentation and
the flavour asymmetry in the proton sea. 
This work  \cite{NSSS}
revises old form factor parametrizations and background contributions 
to one-pion-exchange in comparison with previous 
formulations in order to reproduce 
the new E866 data on the $\bar u-\bar d$ asymmetry. \\
We are aware that the models for the form factors we use here
follow strongly the old approaches and therefore can not reproduce the 
new E866 data as well. 
As we limit our calculations to $z$-values bigger than 0.7, 
the issue of a quantitative description of the $\bar u-\bar d$ asymmetry 
is not so relevant \cite{Kolia}.

In the semi-inclusive reaction we limit
ourselves mainly to forward neutrons in 
photon induced reactions at
small  $x_{\rm Bj}$, as they are studied at HERA. 
In this kinematic region one cannot apply the
usual factorization  into 
a photon-quark cross section times 
a distribution function multiplied by a quark fragmentation function. 
That is why it is interesting by itself to study target
fragmentation as  a nonperturbative process combining
fragmentation and structure information. With good
reason the new concept of fracture functions has been 
invented for this process.
For small $x_{\rm Bj}$ we can consider the
photon as a quark-antiquark state which materializes long 
before it reaches the proton \cite{NZ} and 
interacts with the pion and neutron in the proton wave function. 
Schematically we write the inelastic cross section in a  form similar 
to the proton induced one:
\beqn
\label{gp1step}
\sigma(\phi_0\rightarrow\phi_\alpha) & = & 
        \int\!d^2{\bf b} dw d^2{\bf r}\,|\Psi_{q \bar q }(w,{\bf r})|^2
        \nonumber\\
&&\mbox{} \times \left|\langle\phi_\alpha [1-S^{q\bar q\pi}
        S^{q\bar qn}]|\phi_0\rangle\right|^2 \,.
\eeqn
Here the $q \bar q$ pair wave function is represented by 
$\Psi_{q \bar q }(w,{\bf r})$ with $w$ being the momentum fraction of
the quark and $\bf r$ the transverse separation of the quarks.
This wave function contains the summation over the electric charges
of the different quarks and is electromagnetic in origin. The weak
electromagnetic coupling determines the  cross section of the photon with
the pion. \\
Following the same procedure for the sum over all spatial configurations
and over all excited (inelastic) states we get
\beqn
\frac{d\sigma^{\gamma^* p\rightarrow nX}}{dz} & = & 
        \int\!d^2{\bf b}_{rel}\rho_{n\pi}(z,{\bf b}_{rel})
        \int\!d^2{\bf b} dw d^2{\bf r}
        |\Psi_{q \bar q }(w,{\bf r})|^2 \nonumber\\
&\times & 2Re\Gamma^{q\bar q\pi }({\bf b - s}_{\pi}, {\bf r})
        [1-2Re\Gamma^{q\bar q n}({\bf b - s}_n,{\bf r})]\,,\nonumber\\
&&
\eeqn
where we take into account the dependence on the $q\bar q$-size
by an ${\bf r}$-dependent cross section:
\beq
\label{gammadip}
\Gamma^{q\bar q a}({\bf b},{\bf r}) = \frac{1}{4\pi}\sigma_{\rm
tot}^{q\bar q a}({\bf r})\Lambda_{q\bar q a}^2
\,{\rm exp}\Big[-\frac{{\bf b}^2\Lambda^2_{q\bar q a}}{2}\Big]\,.
\eeq
By performing
the ${\bf b}$-integration analytically  we get
\beqn
\label{desdzgamma}
\frac{d\sigma^{\gamma^* p\rightarrow nX}}{dz} & = & 
        \int\!d^2{\bf b}_{rel}\rho_{n\pi}(z,{\bf b}_{rel})\,
        \int\! dw d^2{\bf r}|\Psi_{q \bar q }(w,{\bf r})|^2
        \nonumber\\
&\times & \sigma_{\rm tot}^{q\bar q\pi}({\bf r})
        \Big\{1-\Lambda_{\rm eff}^2\frac{\sigma_{\rm tot}^{q
        \bar qn}({\bf r})}{2\pi}\,
        {\rm exp}\Big[-\frac{\Lambda_{\rm eff}^2{\bf
        b}_{rel}^2}{2}\Big]\Big\}\,,\nonumber\\
&&
\eeqn
\beq
{\rm with} \hspace{0.5cm} 
\Lambda_{\rm eff}^2 = \frac{\Lambda_{q\bar q\pi}^2\Lambda_{q\bar q
        n}^2}{\Lambda_{q\bar q\pi}^2+\Lambda_{q\bar q n}^2}\,.
\eeq
In this form we see the role played by the photon wave function: 
It governs the integral but does not enter in the magnitude
of the screening correction with the same weight as in the direct term. 
Screening is  a strong
interaction effect which is a function of the transverse
size of the $q \bar q $ pair. 

In the expression \refeq{desdzgamma} we have a 
linear and a quadratic term in the
dipole-hadron cross section $\sigma_{\rm tot}^{q\bar q h}({\bf r})$
averaged on the photon wave function squared. The first is easily
calculated giving the total photon-pion cross section,
\beq
\langle\sigma_{\rm tot}^{q\bar q \pi}\rangle \equiv 
\int\! dw d^2{\bf r}\,|\Psi_{q \bar q }(w,{\bf r})|^2
 \sigma_{\rm tot}^{q\bar q\pi}({\bf r}) = 
\sigma_{\rm tot}^{\gamma^*\pi}\,,
\eeq 
where we have introduced the notation of average to get simpler
expressions in the following.
The second term contains the correlated average of $q \bar q \pi$ and
$q \bar q n$ cross sections.\\
Equation \refeq{desdzgamma} reads 
\beqn
\label{shadeff}
\frac{d\sigma^{\gamma^* p\rightarrow nX}}{dz} &=& \int\!d^2{\bf b}_{rel}
        \rho_{n\pi}(z,{\bf b}_{rel})\,\sigma_{\rm tot}^{\gamma^*\pi^{+}}
        \\
&&\mbox{}\times   \Big\{1-\Lambda_{\rm eff}^2
        \frac{\sigma_{\rm eff}}{2\pi}\,
        {\rm exp}\Big[-\frac{\Lambda_{\rm eff}^2{\bf
        b}_{rel}^2}{2}\Big]\Big\}\,,\nonumber
\eeqn
where we have defined
\beq
\label{sigratio}
\sigma_{\rm eff} = \frac{\langle\sigma_{\rm tot}^{q\bar q\pi}
        \sigma_{\rm tot}^{q\bar qn}\rangle}
        {\langle\sigma_{\rm tot}^{q\bar q\pi}\rangle}.
\eeq 
It is worthwhile to remind here that the total cross sections appearing in
eqs.~\refeq{shadeff} and \refeq{sigratio} depend on  $Q^2$, the photon
virtuality, and on the scaling variables $x_\pi$ and $x_n$ 
(respectively for a pion and a neutron target):
\beq
\label{scalvar}
x_\pi = \frac{x_{\rm Bj}}{1-z} \hspace{.5 cm}
x_n = \frac{x_{\rm Bj}}{z} \hspace{.5 cm} {\rm with\/}\;\;
x_{\rm Bj} = \frac{Q^2}{2q\cdot p} \,.
\eeq
Aware of this we rewrite eq.~\refeq{sigratio} in the following way
\beq
\label{sigratiox}
\sigma_{\rm eff}  =  
        \frac{\langle\sigma_{\rm tot}^{q\bar q\pi}(x_\pi)
        \sigma_{\rm tot}^{q\bar qn}(x_n)\rangle}
        {\langle\sigma_{\rm tot}^{q\bar q\pi}(x_\pi)\rangle}
\, = \, \frac{\langle\sigma_{\rm tot}^{q\bar q n}(x_\pi)
        \sigma_{\rm tot}^{q\bar qn}(x_n)\rangle}
        {\langle\sigma_{\rm tot}^{q\bar q n}(x_\pi)\rangle}\,,
\eeq
where  we use 
$\sigma_{\rm tot}^{q\bar q\pi} \propto \sigma_{\rm tot}^{q\bar qn} $,
keeping
the right $x$- and $Q^2$-dependences.
The quantity in eq.~\refeq{sigratiox} can be then
be parametrized following
Kopeliovich and Povh \cite{KoPo2} as
\beq
\label{sigratioKo}
        \frac{\langle\sigma_{\rm tot}^{q\bar q n}(x_\pi)
        \sigma_{\rm tot}^{q\bar qn}(x_n)\rangle}
        {\langle\sigma_{\rm tot}^{q\bar q n}(x_\pi)\rangle} =
        N_0\frac{1}{F_2^p(x_\pi)}
        \left(\frac{1}{x_\pi}\right)^{\Delta_{\rm eff}}
        \left(\frac{1}{x_n}\right)^{\Delta_{\rm eff}} \,.
\eeq
This parametrization is 
valid in the region of large $Q^2$ and small $x$ values\footnote{Working
at small $x_{\rm Bj}$ we can exchange  all neutron labels
against proton labels.}. 
We choose the
values $N_0 = 2$ GeV$^{-2}$ and $\Delta_{\rm eff} = 0.15$ slightly
different from those quoted in \cite{KoPo2} as we are interested in
a region of smaller $x$ values.
The low value for $\Delta_{\rm eff}$ comes from the fact 
that diffraction and therefore also shadowing
are dominated by processes {\em softer} \cite{NZ} than  
those dominating $F_2^p$ for which we use the 
parametrization in the double leading log approximation 
\cite {KoPo} which covers a wide $x_{\rm Bj}$
and $Q^2$ range.

In the case of real photons at very high energies (HERA kinematics) we 
neglet the small effect in the difference between the photon-pion and
photon-neutron cm-energies and get  
\beqn
\label{sigeff0}
\sigma_{\rm eff}|_{Q^2=0} &=& 
\frac{\langle (\sigma_{\rm tot}^{q\bar q n})^2
        \rangle}{\langle\sigma_{\rm tot}^{q\bar q n}\rangle} 
 =  \frac{16\pi}{\sigma_{\rm tot}^{\gamma p\rightarrow X}}
\left.\frac{d\sigma^{\gamma p\rightarrow pX}}{dt}\right|_{t=0}\nonumber\\
& = &  16\pi\frac{b_D\sigma_D}
        {\sigma_{\rm tot}^{\gamma p\rightarrow  X}}
\approx  20\, {\rm mb}\,,
\eeqn
where we use $d\sigma^{\gamma p\rightarrow pX}/dt 
= b_D\sigma_D\,$exp$(b_D t)$ with 
the experimental values for $b_D$ and $\sigma_D$  according to
\cite{Zediff,H1diff}. 

Finally following the same idea of shadowing as a soft process which
motivated the choice of $\Delta_{\rm eff}$ 
we employ slope parameters $\Lambda^2_{q\bar q n}$ 
and $\Lambda^2_{q\bar q \pi}$
which are calculated from total cross sections\footnote{From the neutron
target to the pion target we scale 
total cross sections with the ratio of pion to proton size,
i.e.  with a factor 2/3.}
${\sigma_{\rm tot}^{q\bar q n}} = \sigma_{\rm tot}^{\rho n}\approx 30$ mb 
and ${\sigma_{\rm tot}^{q\bar q \pi}} \approx 20$~mb:
\[ 
\Lambda^2_{q\bar q \pi} \approx \frac{4\pi}{\sigma_{\rm tot}^{q\bar q
\pi}}
\hspace{1cm}
\Lambda^2_{q\bar q n} \approx \frac{4\pi}{\sigma_{\rm tot}^{q\bar q n}}
\,.\]
For the double differential $\gamma ^* p$ cross section, 
we obtain with the help of eqs.~\refeq{d3csgl} and \refeq{normflux},
\beq
\frac{d^2\sigma^{\gamma^* p\rightarrow nX}}{dzdp_t^2} 
 = \frac{1}{N(z)}\frac{dN(z,p_t)}{dp_t^2}\,
\frac{d\sigma^{\gamma^* p\rightarrow nX}}{dz} \,,
\eeq
which is then rescaled to yield the differential cross section 
for neutron production in electron-proton scattering 
(besides a small longitudinal contribution)
\beq
\frac{d^4\sigma^{e p\rightarrow e'nX}}{dx_{\rm Bj}dQ^2dzdp_t^2} = 
\frac{\alpha_{\rm em}}{2\pi x_{\rm Bj}Q^2}\,(2-2y+2y^2)\,
\frac{d^2\sigma^{\gamma^* p\rightarrow nX}}{dzdp_t^2}\, ,
\eeq
with $ y = \frac{Q^2}{x_{\rm Bj}s}$; 
$s$ is the cm $ep$ total energy squared.\\
The phenomenological $\gamma^*\pi^+$ cross section entering 
eq.\refeq{shadeff} 
is expressed in terms of the pion structure function  
\beq
\label{F2pi}
\sigma^{\gamma^*\pi^+}_{\rm tot}(x_\pi,Q^2) = 
\frac{4\pi^2\alpha_{\rm em}}{Q^2}F_2^{\pi^+}(x_\pi,Q^2)\,.
\eeq

In the following we use for demonstration purpose\\  
$F_2^{\pi^+}(x_{\rm Bj},Q^2)~=~\frac{2}{3}F_2^p(x_{\rm Bj},Q^2)$ 
valid at $x_{\rm Bj}\ll 0.1$. 
For real photons we refer to the fit performed in \cite{DoLa}.

\begin{figure}[!tb]
\center
\mbox{\epsfig{file=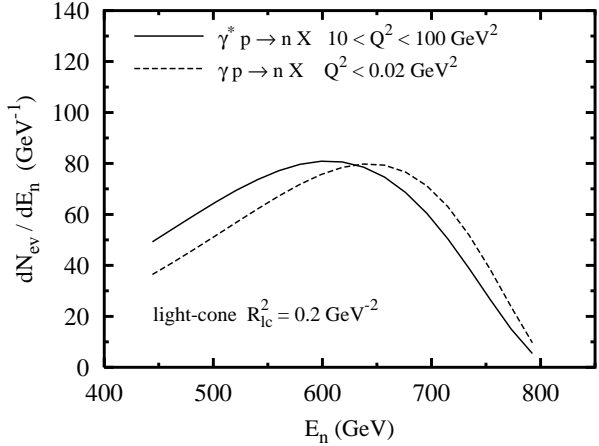,width=8.5cm}} 
\caption{\label{photodis} 
{Comparison of the energy distributions of neutrons for deep inelastic
events and photoproduction imposing the kinematical cuts of
ref.~\cite{Ze}. 
The distributions are normalized to the total number of DIS events.
}}
\end{figure}

Averaging the colour dipole-neutron cross section together 
with the  dipole-pion one, 
the decrease of the effective cross section $\sigma_{\rm eff}$ 
is less steep
with increasing the virtuality of the 
photon than e.g. the free dipole-neutron cross section. 
Still screening is reduced for high $Q^2$. 
To show
this we plot in fig.~\ref{photodis}  the integrated neutron energy
distributions in photoproduction 
with $Q^2< 0.02$ GeV$^2$, 
5 GeV $ <E_e<$ 22 GeV, 
 and in deep inelastic scattering
with 10 GeV$^2$ $< Q^2 < 100$ GeV$^2$,
$0.04<y<0.95$,   the ZEUS
cuts ($\theta_{\rm scat}<0.6$ mrad, $p_t < 0.5$ GeV and  
an integrated luminosity of 6.7 pb$^{-1}$).  
A shift of the DIS
cross section maximum by 50 GeV to lower energies
is clearly visible. This is due to
screening, which reduces the cross section mainly at 
smaller $E_n$ (neutron energy), 
such that the peak appears at higher energies in
photoproduction.
Our results indicate a clear signal for more transparency of the
neutron when interacting with highly virtual photons. 
Small size color dipoles rescatter less on the target fragment neutron, 
even if for very large $Q^2$ screening still persists. The 
semi-inclusive reaction with Regge-exchange also allows
to study the same phenomena as in  the 
diffractive $\sigma^{\gamma^* p \rightarrow p+ X}$ cross section.

\begin{figure}[thb]
\center
\mbox{\epsfig{file=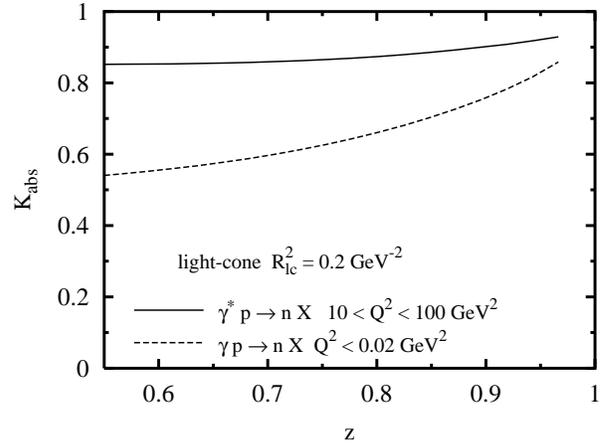,width=8.5cm}} 
\caption{\label{kabsgp} 
{ Absorptive $K$-factor for neutron production in real and
virtual photon-proton collisions.
}}
\end{figure}

The absorptive $K$-factors for $\gamma p\rightarrow 
nX$ and $ep\rightarrow e'nX$ are shown in fig.~\ref{kabsgp}: it is visible
that for DIS neutron production (high $Q^2$)
screening becomes very weak , while for real photons we get, 
at large $z$, a somewhat reduced effect 
compared to proton-proton collisions (see fig.~\ref{kabspp}). 
For $z > 0.75$ and $Q^2 > 10$ GeV$^2$ 
the factorization of the cross section into a pion
flux factor and a pion structure functions looks very acceptable.

\begin{figure}[!thb]
\center
\mbox{\epsfig{file=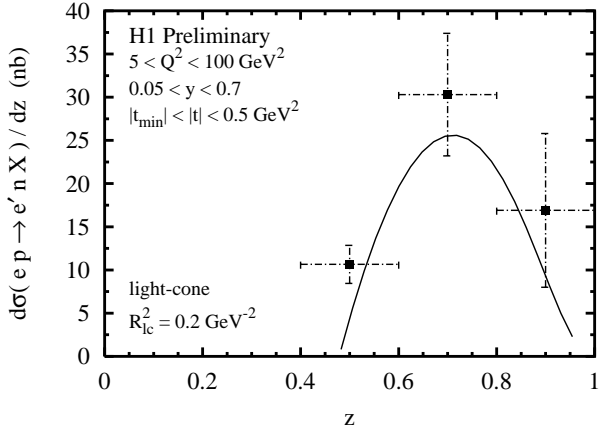,width=8cm}} 
\caption{\label{desH1} 
{ Comparison of the differential cross section including screening
corrections in light-cone approach with preliminary H1 data, 
ref.~\cite{H1}. 
}}
\end{figure}

In fig.~\ref{desH1}  we show the $p_t$-integrated neutron cross
section with
the H1 cuts as function of $z$.
The screening in $pp \rightarrow nX $ manifests itself as
a bigger $ep \rightarrow e'nX$ cross section, since a higher pion flux 
is needed to explain the
measured hadronic cross section. 
Notwithstanding the validity of the impuls approximation
in  high-$Q^2$ processes, factorization is broken in  soft hadronic 
fragmentation reactions. The comparison of our result with the preliminary
H1 data is very encouraging.

The fracture functions \cite {TrVe,Gr} introduced to describe
target fragmentation allow a model independent evolution in
the framework of perturbative QCD, albeit the starting functions have to 
be known. In this approach the  nonperturbative distributions are
related to the
$p_t$-integrated differential cross section for leading neutron production
in $ep$ scattering in the following way
\beq
\frac{d^3\sigma^{e p\rightarrow e'nX}}{dx_{\rm Bj}dQ^2dz} =
\frac{2\pi\alpha^2_{\rm em}}{ x_{\rm Bj}Q^4}(2-2y+2y^2)\,
M_2(x_{\rm Bj},Q^2,z)\,,
\eeq
where at leading order in $\alpha_s$ there is no contribution from the
fragmentation of the struck quark in the target region.
In the one-pion-exchange model without screening corrections one can
interprete\footnote{Here one should consider this fracture function as an
input distribution at
a certain scale $Q_0^2$ and then apply the pQCD-evolution.}  
 $M_2(x_{\rm Bj},Q^2,z)$ as the product of the flux of
neutrons integrated over $p_t$, times the pion structure
function $F_2^\pi(x_\pi,Q^2)$.\\
The fracture function $M_2$ is defined in analogy with
the structure function $F_2$, 
\beq
M_2(x,Q^2,z) = x\sum_i e_i^2 M_{i,n/p}(x,Q^2,z)\,,
\eeq
where $M_{i,n/p}(x,Q^2,z)$ represents the probability of finding
a parton of flavour $i$ with momentum fraction $x$ and a neutron
with momentum fraction $z$ inside a proton.

\begin{figure*}[!bt]
\center
\mbox{\epsfig{file=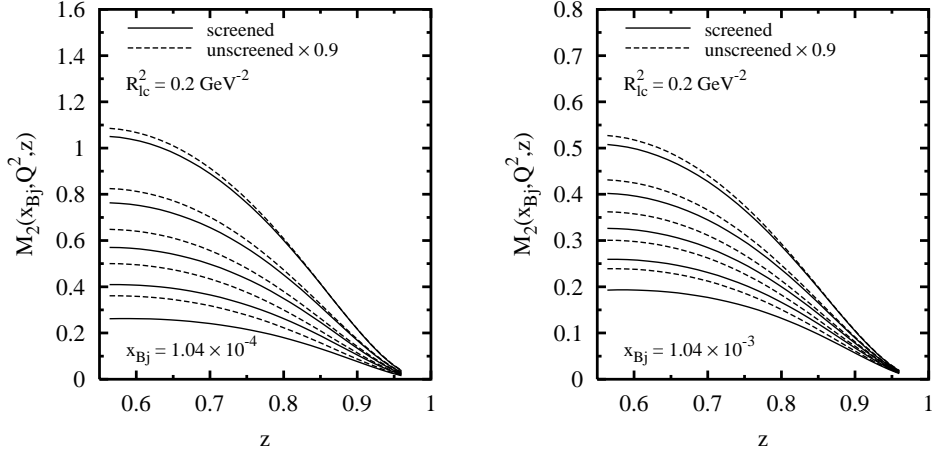,width=13cm}} 
\caption{\label{fract} 
{ $Q^2$-evolution of fracture functions
$M_2(x_{\rm Bj},Q^2,z)$ at fixed
$x_{\rm Bj}$ as function of $z$. The $Q^2$ values are (from 
bottom to top) 2.5, 4.4, 7.5, 13.3, 28.6  GeV$\,^2$. Unscreened
results are scaled by 0.9.
}}
\end{figure*}

In  fig.~\ref{fract} we give the fracture functions
$M_2(x_{\rm Bj},Q^2,z)$ as functions of $z$, at fixed $x_{\rm Bj}$,
for various $Q^2$ values. 

The $Q^2$-evolution of the
fracture functions we use is given at small virtualities by
the 
higher twist effect of screening and at higher $Q^2$ by the evolution 
of the pion structure function. In fact for neutron fragmentation 
the Altarelli-Parisi evolution of the fracture function dominates
the total $Q^2$-dependence as has been
argued in ref.~\cite {FlSa}.

\section{Discussion of fragmentation results and 
validity of the factorization hypothesis}

Finally we compare the screening corrections 
for the three
different cases of proton, real and DIS photon induced
semi-inclusive fragmentation reactions (cf. figs.~\ref{kabspp},
\ref{kabsgp}). 
One sees that both proton induced and real
photon induced cross sections have $K$-factors differing by about 
thirty percent from unity for $z<0.8-0.9$.
In this region factorization for these reactions does not hold.  
On the other hand for deep inelastic scattering at $Q^2 > 10$ 
GeV$^2$ the rescattering of the $q \bar q$ dipole in the photon by the
neutron is weaker. 
The different sizes of the absorptive corrections in $pp$ compared to 
$\gamma^* p$ interactions prevent a model independent extraction of 
$F_2^\pi$ from the simple factorization hypothesis,  in agreement 
with the conclusions reached in ref. \cite{NSZ}. 
We also show that the effect of rescattering for real photon-proton 
reactions is important and that one can 
disentangle the difference with highly virtual photons:
the shift in the neutron energy distributions 
can be understood in terms of different absorption effects.

In addition an extraction of the
pion structure function 
is affected by the uncertainty in the
determination of the pion flux factor. Our fits to the 
proton induced fragmentation reaction give for the radius parameters 
$R_{lc}^2= 0.2\, {\rm GeV}^{-2}$ and  $R_c=0.05\,{\rm GeV}^{-2}$. 
More accurate $p_t$ neutron spectra  in the case of deep
inelastic scattering could reduce this uncertainty significantly. At the
moment the $p_t$ spectra in proton induced reactions have been  measured
for too low energies \cite {Bl} or too high $p_t$ in the relevant $z$
range \cite {En}.

It may
be added that  other mechanisms in all three reactions start
contributing for $z<0.75$; $\rho$-exchange \cite {HLNSS,KPP,NSSS} 
has been added to the 
strong proton induced cross section
presenting a non negligible 
contribution for $0.5<z<0.7$. 
In Monte Carlo simulations (LEPTO) the fragmentation
cross section increases for  $z<0.5$.
Further work is needed to understand this part of the
cross section theoretically. 

The semi-inclusive reactions in the large $z$-region
show a case of transparency reminiscent of diffractive
events. We think 
this finding of our paper merits a more accurate experimental 
examination. Exclusive (e.g. $\rho$-production )
reactions are in seemingly good agreement with theoretical
calculations based on the dipole picture, which 
is underlying color transparency.
What is important to realize is that the 
final state interactions of virtual partons
appear correlated with their initial state interaction. 

\subsubsection*{Acknowledegments}
\begin{acknowledgement}
We acknowledge many valuable discussions with
J.~H\"ufner, B.Z.~Kopeliovich, B.~Povh, A.~Sch\"afer, D.~Jansen
and T.~Nunnemann. Special thanks go to N. N. Nikolaev for his
kind help in revising the manuscript and pointing out the important
reference \cite {NSSS}.

U.~D'Alesio was funded through the European TMR Contract 
No.~FMRX-CT96-0008: Hadronic Physics with High Energy Electromagnetic
Probes.
\end{acknowledgement}


\appendix

\section*{Appendix}

Here we give the expression for the function entering  the
Skyrme-type form factor, eq.~\refeq{Sky}, \cite{Fr}:
\beq
g(x) = 1 + \frac{\sum_{k=0}^3 a_k T_k(\bar x)}{1+mx}\,, \hspace{.5cm}
x  =  \frac{|t|}{m_\pi^2}\,,\hspace{.5cm} \bar x = \frac{2x-a-b}{b-a}
\eeq
where 
$T_k$ is the k-th Tchebycheff polynomial of 1st kind, with 
$
      a  =  0.003113\;\;\;
      b  =  280.198488\;\;\;
      m  =  10^{-8}\\
    a_0  =  6.242336\;\;\;
    a_1  =  4.940353\;\;\;
    a_2  =  2.740654\;\;\;\\
    a_3  =  0.9217577 \;.
$


\end{document}